\documentclass[runningheads]{llncs}

\usepackage{graphicx}
\usepackage{hyperref}

\usepackage{orcidlink}

\begin{document}

\title{Visualisation to Explain Personal\\ Health Trends in Smart Homes}

\author{Glenn Forbes \orcidlink{0000-0003-0282-3089} \and
Stewart Massie \orcidlink{0000-0002-5278-4009} \and
Susan Craw \orcidlink{0000-0003-1870-0323}
$\left\{g.r.forbes, s.massie, s.craw \right\}$@rgu.ac.uk
}

\authorrunning{G. Forbes et al.}

\institute{Robert Gordon University, Aberdeen, UK}
\maketitle

\begin{abstract}
 An ambient sensor network is installed in Smart Homes to identify low-level events taking place by residents, which are then analysed to generate a profile of activities of daily living. These  profiles are compared to both the resident’s typical profile and to known “risky” profiles to support recommendation of evidence-based interventions. Maintaining trust presents an XAI challenge because the recommendations are not easily interpretable. Trust in the system can be improved by making the decision-making process more transparent. We propose a visualisation workflow which presents the data in clear, colour-coded graphs.

\end{abstract}

\section{Introduction}

Artificial Intelligence is a powerful technology with many potential applications in healthcare. IoT sensor networks combined with long term data collection are starting to be adopted in Smart Homes as a new approach for health monitoring. As with many of the AI applications used for decision making, Smart Home health monitoring systems are black box systems that give little explanation for why specific interventions are suggested. Insights generated from these systems are often unclear or impossible for humans to understand and so introduce the danger of reducing user trust in a system which aims to help. One approach to inspiring confidence and trust is to improve transparency in the decision-making process so that it is explainable to stakeholders, including residents, family members and carers.

Many countries are facing an ageing population which puts additional strains on health and social services with a smaller proportion of working population available to support services. In this changing scenario it is important to help people with mobility or social needs to live independently for longer, and so reduce their reliance on more expensive health care solutions. We examine our solution for Smart Homes to support assisted living environments by identifying and reporting on resident health trends. Sensors are installed in new-build Smart Homes with the aim of supporting tenants to live independently for longer. Specifically, in this paper, we address one of the main challenges of explanation by visualising the decision process of a real-world health monitoring system that senses and predicts the level of risk of falling attributed to Smart Home residents. These predictions can then be relayed through dashboards using simple graphs.

\section{Related Work}

Activities of Daily Living (ADLs) are events in daily life which would be considered intrinsic to a person's ability to live independently. ADLs include being able to dress, get out of bed, and feed oneself. Katz~\cite{Katz1963} originally proposed the term along with a scale for rating a person's independent ability using their performance in simple ADLs. The concept of losing ADLs as we age has influenced future research in the field by identifying that specific ADLs are more indicative of reduced capability than others. IoT offers an opportunity to provide continuous behavioural and physiological monitoring of residents. 

Once activities are being regularly tracked in a home environment, the resident's routine behaviours emerge~\cite{Aipperspach2010}. Changes in these routines compared to the resident's historical performances can be indicative of encroaching health problems. Ongoing aberrant behaviour may require healthcare interventions. However the direct link between the resident's behaviours and the relevant interventions may not always be immediately clear to an inexperienced observer~\cite{panigutti2020doctor}. Visualisations can be used to aid stakeholders in interpreting the complex sensor data insights \cite{massie2004visualisation}.

\section{Smart Home Activity Profiles}

FitHomes is an initiative, lead by Albyn Housing Society Ltd to support independent living with the supply of custom-built Smart Homes fitted with integrated non-invasive sensors. Sixteen houses have been built at Alness near Inverness, Scotland, with an additional eight retrofit houses in Nairn. The initial project focused on using sensor data to develop a prototype fall prediction system for the residents of FitHomes, and now aims to monitor a wider range of long term health conditions over time. One aspect is to identify increased risk of falls and so a key focus for monitoring is to identify activity levels, patterns, and speeds. However, monitoring can go beyond just movement to consider other factors that have been shown to be related to falls, including  dehydration, tiredness, and mental health. Gaining information on these additional factors requires monitoring to also capture data on more general activities such as eating \& drinking behaviours, sleep patterns, and toileting \& grooming habits. With these criteria in mind a range of sensors have been selected  including: IR motion sensors; contact sensors to capture room, cupboard, and fridge door opening; pressure sensors for bed \& chairs; electricity smart meters to identify power usage pattern; float sensors identifying toilet flushing; and humidity sensors to identify shower use.

\begin{figure*}[!ht]
	\centering
	\includegraphics[width=0.8\textwidth,keepaspectratio]{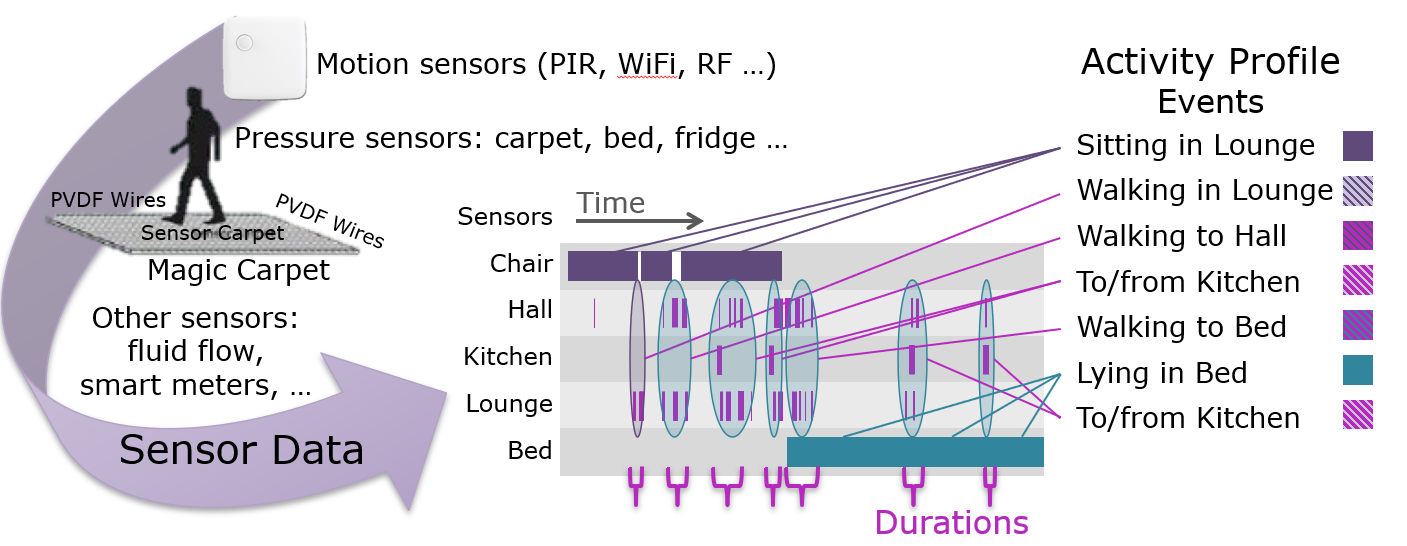}
	\caption{Identifying activities from sensor activations.}
	\label{fig:adls}
\end{figure*}

Most of the sensors chosen have a binary output which simply activates when the event they are monitoring takes place e.g. a door opening; however others output continuous readings provided at fixed polling rates. The data fusion task across multiple sensors with different output modes becomes one of the main challenges. We employ a two stage process: the first is to generate activity profiles, see Figure~\ref{fig:adls}, and uses pattern identification to create meaningful representations from the raw sensor data that captures the residents’ activities e.g. sleeping, showering, cooking; and then the second stage is to assemble these activities into personalised daily profiles. The second stage is the analysis of these profiles, see Figure~\ref{fig:cbr}. Changes in the resident’s own activity patterns can then be used to detect deterioration in health, while comparisons with the patterns of other Smart Home residents can provide benchmark measures. The data thus supports evidence-driven intervention tailored to the resident and their specific circumstances. 

\section{Reasoning with ADLs}

Identifying ADLs in themselves does not give an indication of health. However, it has been shown that functional assessment is an effective way to evaluate the health status of older adults~\cite{Cook2015}; ADLs are lost as we age, and FITsense monitors changes in ADL activity as an indicator of deteriorating health and increased risk of falls. A CBR approach is adopted. In our scenario, a set of ADL templates is combined with contextual information to form the problem representation to retrieve similar profiles from a case base.
%Solutions will identify interventions, where required, and their previous outcomes. 

\begin{figure*}[!h]
	\centering
	\includegraphics[width=0.8\textwidth,keepaspectratio]{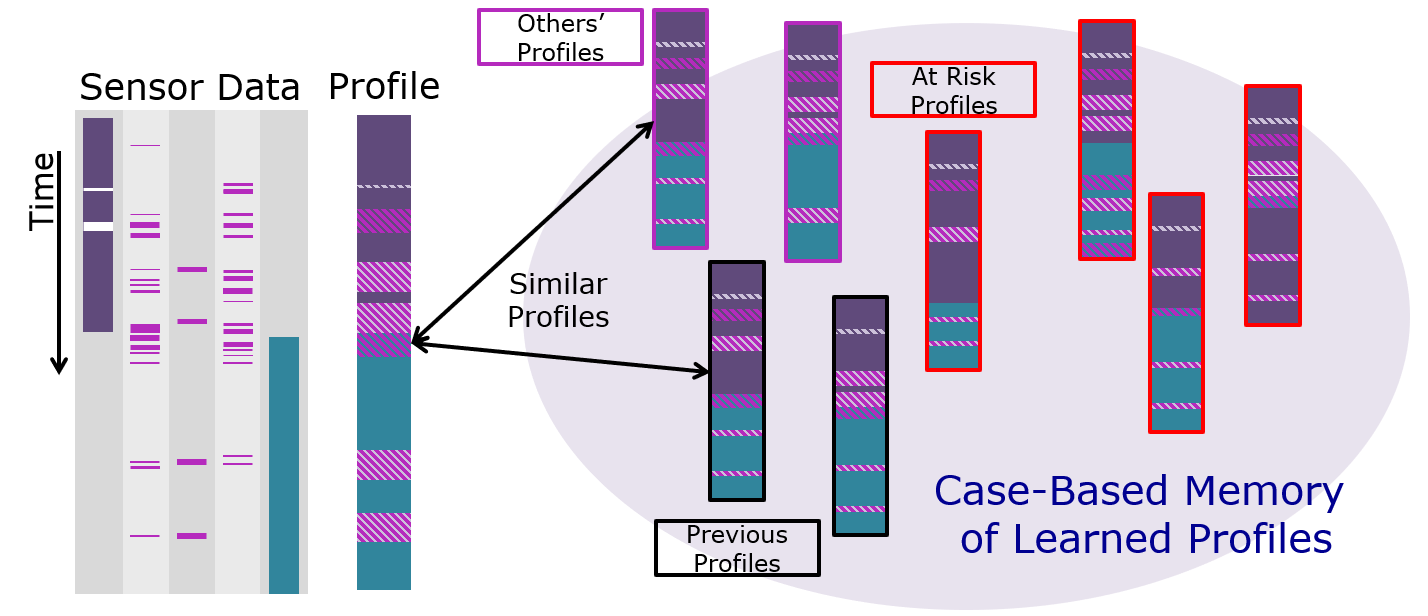}
	\caption{CBR Approach to Identifying 'Risky' Behaviours}
	\label{fig:cbr}
	\centering
\end{figure*}

Figure~\ref{fig:cbr} presents an overview of our approach. Low-level, time-stamped events identified by the sensors are transformed into a daily user profile. The profiles are based on a set of features extracted from ADL sequences, with mixed data types: some features are binary, e.g. disturbed sleep; some features are counts, e.g. number of room transitions or stand up from seat count; some are cumulative daily time spans, e.g. time sitting; while others are numeric, e.g. average gait speed. Whatever the data type a similarity measure is associated with each feature so that comparison can be made between them. A set of daily ADL profiles for a resident can then be compared with those in the case base, on the right of Figure~\ref{fig:cbr}. Retrieval of similar profiles labelled as at risk identifies the need to recommend intervention, and falling similarity with the user's own previous profiles identifies changing behaviours. Importance in determining similarity for FITsense is given to features known to correlate with falls. For other health conditions the similarity knowledge could be refined to reflect specific conditions e.g. repetitive behaviour for dementia, general physical activity level for obesity, etc.

\section{Explanation}

Case-based Reasoning systems can facilitate comparisons between cases, which is important to explanation because they are real examples. However, in this form their representation is complex, formed as a time series of behaviours. In order for users to establish meaning from this data, they have to manually interpret it and interrogate differences. While this may be a viable option for healthcare professionals, it is both not easily accessible, or an efficient use of time. Furthermore, this approach does not consider explanation for whether an intervention would be necessary. To consider explanation both for behavioural observations and recommended healthcare interventions, we need to present a high-level summary of aspects of behaviour which could be indicative of fall risk. %to encourage residents/carers/family to trust the recommended interventions.

Six main features were selected for fall risk determination: Sleep, Sleep Disturbances, Room Transitions, Activity, Wandering, Toilet. These features being identified by investigation of fall risk indicators, which are typically highlighted through GP and hospital testing \cite{forbes2020fall}. Each feature is extracted from profiles by identifying count and duration of specific ADL performances or sequences, from which linear scores can be calculated. Factor scores are generated using an upper and lower threshold initially established using expert domain knowledge and can be refined as additional data is collected.

\begin{figure*}[!h]
	\centering
	\includegraphics[width=0.6\textwidth]{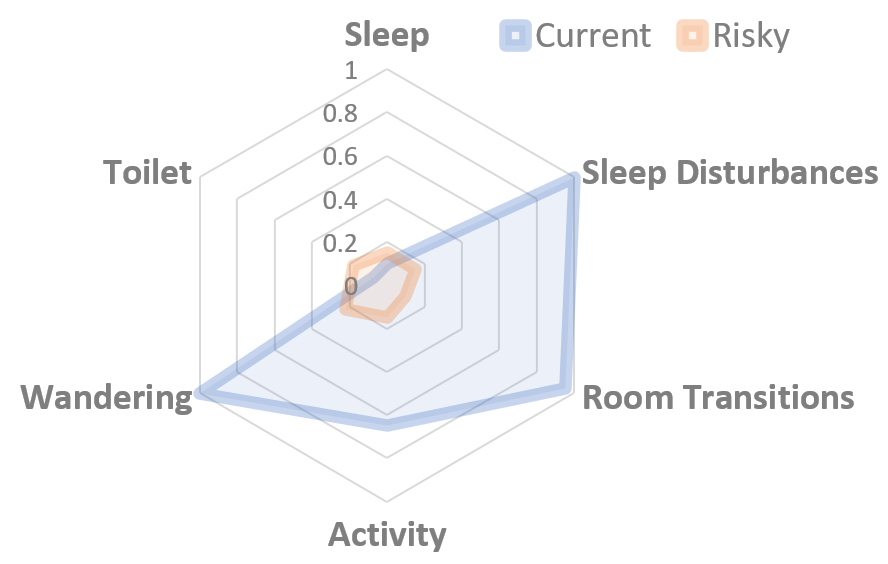}
	\caption{Radar Chart visualisation of a resident's daily fall risk profile.}
	\label{fig:radar}
	\centering
\end{figure*}

We present a radar chart visualisation of a resident's fall risk profile in Figure~\ref{fig:radar}. By linking each of the profile features, we generate a summary as a snapshot of the current fall risk factors of the resident. This can be compared to historical profiles from the same or similar residents, or to a risky profile. A prototypical risky profile example is displayed in the figure, which was generated through consultation with healthcare professionals. As the system progresses risky profiles will be generated using personalised thresholds. In the example, we can see the resident could be considered as at risk based on their Toilet and Sleep scores. This shows that there are issues with the resident's behaviour which need to be addressed, such as low quality sleep and over-frequent toilet use. These profiles can also be compared over time to view long term trends in resident health factors.

\section{Conclusions}

AI systems for behavioural-based health monitoring can produce meaningful insights, however many are considered faceless, black-box systems. We have developed an explanation and accompanying visualisations which make output from a complex system more transparent and accessible to stakeholders. This should increase confidence and trust in our AI healthcare systems.

In future work, we aim to establish feature sets for a wider range of long-term health conditions, such as dementia and cancer. We also plan to continue reworking and refining the existing feature set to ensure robustness as the system learns from ongoing data collection.

\subsubsection{Acknowledgements}
This work was funded by Albyn Housing Society, and the Scottish Funding Council via The Data Lab innovation centre.

\bibliographystyle{splncs04}
\bibliography{library}

\end{document}